\documentclass[reprint,apd,superscriptaddress,showkeys,nofootinbib,longbibliography]{revtex4-1}

\usepackage{amsmath}
\usepackage{amsfonts}
\usepackage{graphicx}
\usepackage[breaklinks=true,colorlinks=true,linkcolor=blue,urlcolor=blue,citecolor=blue]{hyperref}
\usepackage{bm}
\allowdisplaybreaks

\begin{document}

\title[Efficient Estimation of Barycentered Relative Time Delays]{Efficient Estimation of Barycentered Relative Time Delays for Distant Gravitational Wave Sources}

\author{Orion Sauter}
\email{osauter@umich.edu}
\affiliation{University of Michigan, 450 Church Street, Ann Arbor, Michigan 48109, USA}

\author{Vladimir Dergachev}
\email{vladimir.dergachev@aei.mpg.de}
\affiliation{Max Planck Institute for Gravitational Physics (Albert Einstein Institute), Callinstrasse 38, 30167 Hannover, Germany}
\affiliation{Leibniz Universit\"at Hannover, D-30167 Hannover, Germany}

\author{Keith Riles}
\email{kriles@umich.edu}
\affiliation{University of Michigan, 450 Church Street, Ann Arbor, Michigan 48109, USA}

\begin{abstract}
Accurate determination of gravitational wave source parameters relies on transforming between the source and detector frames.  All-sky searches for continuous wave sources are computationally expensive, in part, because of barycentering transformation of time delays to a solar system frame. This expense is exacerbated by the complicated modulation induced in signal templates. We investigate approximations for determining time delays of signals received by a gravitational wave detector with respect to the solar system barycenter. A highly non-linear conventional computation is transformed into one that has a pure linear sum in its innermost loop. We discuss application of these results to determination of the maximal useful integration time of continuous wave searches.

\end{abstract}

\keywords{LIGO; Gravitational Waves; Solar System Modeling}

\maketitle

\section{Introduction}
The hunt for the first detection of continuous gravitational waves (CW) is under way with many searches published \cite{CasAPaper,CrabPaper,EarlyS5PowerFluxPaper,FullS5PowerFluxPaper,houghS2,LSC:05-S2TD,LSC:07-S34TD,O1knownPulsars,orionspur,PhysRevD.93.042007,PhysRevD.94.064061,PhysRevD.95.082005,PhysRevD.96.062002,PhysRevD.96.122004,PhysRevLett.118.121102,S1PulPaper,S2Fstat,S4EaHpaper,S4IncoherentPaper,S5Hough,S5R1EaHpaper,S5R5EaHpaper,S5Targeted,S6Bucket,S6PostCasAPaper,S6PowerFluxPaper,S6SpotlightPaper,VSR1fstat} or in progress. Some searches target known potential sources, such as the Crab Pulsar, but other searches look for unknown sources over a broad frequency band and cover large parameter spaces.  These algorithms make use of substantial computational resources, so any reduction in computational demands is helpful.  In this paper, we examine one aspect of the searches amenable to simplification: calculation of the time delays of signals received by a gravitational wave detector with respect to the solar system barycenter for an ensemble of assumed source sky positions.

The barycentric time corrections play an important part in the signal-space geometry because the signal templates have the general form $A(t, p)\exp(i \Phi(t, p))$, where amplitude terms $A(t,p)$ are slowly varying, but the phase $\Phi(t, p)$ can vary rapidly. Accurate determination of the phase relies on precise positioning of the detectors in space and time, as this dictates the signal arrival time. Thus most of the influence of sky position mismatch comes through barycentric corrections that act mostly on phase.

We investigate approximate models for time delays, characterized here by ``emission time'', the inferred time of signal emission in the solar system barycenter frame for a given signal reception time at the detector and the distance to the source. In addition, our semi-analytic formula provides an efficient way to compute sets of barycentric corrections for nearby templates using piece-wise polynomial approximations. The analysis of barycenter timing corrections for the Earth-Sun system serves in addition as a model for a general circularized binary with small modulation depth. As the modulations of source and detector add independently, such an analysis could, in principle, be applied to a binary source simply by doubling the number of terms for an assumed signal model.

A recently published paper \cite{ROMpaper} explored reduced order modeling with respect to time of barycentering for a targeted search of a single sky location. To simplify computation the authors of the paper did not model Shapiro delays which add considerable non-linearity to emission time correction. Our approach addresses spatial dependence of barycentering for discrete time intervals, including all terms, in particular Shapiro delays, and produces models with corrections smaller than any practical tolerances in the wide-parameter searches. We specify our basis vectors explicitly, in terms of well-understood time periods and explicit sky-position dependent functions. The decomposition is performed with factor analysis. Our ability to tackle highly-nonlinear corrections is largely due to modeling of differential emission times.

Traditional computation of emission times involves trigonometric functions, square roots, and divisions which are very expensive operations on any modern hardware. The decomposition presented in this paper transforms this computation into a bilinear product of a few hundred precomputed terms. This computation is performed using only addition and multiplication and is easily vectorized. While our algorithm involves fewer operations than traditional computation, the fact that no trigonometric or other special functions are used places it in a class of its own, with speed not dependent on the efficiency of system libraries, resulting in speedup of barycentering calculations of potentially several orders of magnitude for applications to all-sky searches.

While this improved computational efficiency is most welcome, the most expensive all-sky broadband searches have inner loops that iterate over frequency and spindown, and the cost of computing barycentric corrections is largely amortized away. In these situations the algorithm's greater contribution is enabling {\em Loosely Coherent} methods. The {\em Loosely Coherent} algorithms \cite{pfloose, pfloose2, loosely_coherent3} used in recent CW searches \cite{s5,s6,o1allsky} are constructed to process sets of signal templates close in sky position, spindown and frequency. In the limit of infinitely dense placement these template sets form a manifold, the geometry of which influences the efficiency of the algorithm and, of course, its technical implementation. The model discussed here provides explicit expressions for the type of degeneracies exploited by the {\em Loosely Coherent} algorithms.

Lastly, the explicit identification of basis vectors, in particular the dependence of phase on sky position mismatch, permits us to show that fully coherent single-interferometer large-parameter-space searches are not optimal. Rather, a semi-coherent search maximizes detection efficiency, and increasing integration time further can actually decrease detection efficiency.

Even for analysis of data from multiple interferometers this result still holds in cases for which there are additional search parameters, such as  frequency derivatives. For example, when data from two interferometers is analyzed the dimension of the fully coherent data considered as a vector space over real numbers is four. Any search iterating over initial signal phase, right ascension, declination, and first frequency derivative (spindown) will have four independent parameters to fit to the fully coherent data. In other words, the effective bandwidth of the signal is larger than is naively assumed in a fully coherent search.

In the following, section~\ref{sec:mathmodel} summarizes the exact mathematical model we use for barycentering and the motivation for an approximate version.
Section~\ref{sec:implementation} outlines the structure of a practical implementation of the approximation procedure.
Section~\ref{sec:example} describes in detail a particular demonstration example of an approximation implementation.
Section~\ref{sec:results} summarizes results from applying the example implementation to the time span of the first
Advanced LIGO data run.
Section~\ref{sec:conclusions} discusses the implications of the results.

\section{Mathematical model}
\label{sec:mathmodel}

Precise barycentering has been important to pulsar astronomy for decades.
A widely used expression for emission time in the pulsar frame ($t^\mathrm{psr}_\mathrm{e}$) is given by Edwards, \emph{et al.} as \cite[Eq.~7]{tempo2}
\begin{equation}
t^\mathrm{psr}_\mathrm{e} = t^\mathrm{obs}_\mathrm{a} - \Delta_\odot - \Delta_\mathrm{IS} - \Delta_\mathrm{B},
\end{equation}
where $t^\mathrm{obs}_\mathrm{a}$ is the arrival time at the observatory, $\Delta_\odot$ is the time delay from transforming from the detector frame to the SSB, $\Delta_\mathrm{IS}$ is the travel time in the interstellar medium, and $\Delta_\mathrm{B}$ includes transformation to the pulsar frame for binary systems.

We reframe this in terms of searches for continuous gravitational waves:
\begin{equation}
T(t, u, p) = t - \Delta_\odot(t, u) - \Delta_\mathrm{IS} - \Delta_\mathrm{B}(t, p).
\end{equation}
The emission time $T$ is a function of detector local time $t$, source location $u$ and intrinsic source parameters $p$ (for a source in motion). Because modern computer architectures are vector-based, it is typically more efficient to compute arrays of values of $T(t, u, p)$ for sets of times $\mathcal{T}=\left\{t_i\right\}$ and templates $S=\left\{(u_j,p_j)\right\}$.

For a single template $(u_0,p_0)$ the function $T(t, u_0, p_0)$ has a very non-trivial behaviour due to several nearly periodic influences from the Sun, planets, and the Moon as well as contributions from General Relativity. 

Because any analysis method must overlap templates $(u, p)$ closely enough to provide sufficient detection coverage, we can expect to compute 
arrays $T(\mathcal{T}, u, p)$ for nearby $(u,p)$. 

Therefore, we separate the problem into two parts: computation of $T(\mathcal{T}, u_0, p_0)$ for a fixed template $(u_0, p_0)$; and
computation of differences $\Delta(\mathcal{T}, u, p; u_0, p_0)=T(\mathcal{T}, u, p)-T(\mathcal{T}, u_0, p_0)$. When sets $\mathcal{T}$ and $S$ are finite the isomorphism of vector spaces ${\mathbb R}[T \times S]$ and ${\mathbb R}[T] \otimes {\mathbb R}[S]$ implies there exists the following decomposition:
\begin{equation}
\label{decomp_equation}
\Delta(t_i, u_j, p_j)=\sum_{k=1}^N f_k(t_i) g_k(u_j, p_j)
\end{equation}
where $f_k(t_i)$ and $g_k(u_j, p_j)$ are, in general, arbitrary single-valued functions. Such decompositions in more general situations such as continuous or algebraic functions have been studied extensively. The work goes back to the 13$^{\textrm{th}}$ problem by Hilbert, with one of the main results being the Kolmogorov-Arnold representation theorem \cite{kolmogorov, arnold}. These decompositions are often used for data compression and work well even in the case of very wideband signals such as compact binary coalescences \cite{saturated}.

The key to our approach is that it is possible to find an approximate version of Equation \ref{decomp_equation} with a number of terms $N$ much smaller than the dimensionality of space spanned by $\Delta(t_i, u_j, p_j)$. The well-understood equations of motion of the Solar system allow us to use explicit time and space dependent factors and perform a simple linear regression to find the coefficients. 

Besides providing computational efficiency this analysis identifies analytical functions $f_k$ and $g_k$, paving the way for developing advanced {\em Loosely Coherent} \cite{pfloose, pfloose2, loosely_coherent3} semi-analytic statistics.

\section{Practical implementation}
\label{sec:implementation}
%\begin{figure*}[htbp]
% \centering
% \includegraphics[width=\textwidth]{diffMaps}
% \caption{\label{fig:rotationMaps} Example of emission time variation with sky position. The middle panel shows differences between computed emission time and a reference time. The top panel shows differences in emission times between points offset in declination by $0.01$\,rad, while the bottom panel shows differences in emission times between points offset in right ascension by $0.01$\,rad. Note that the color scales for difference plots are greatly reduced.
% }
%\end{figure*}

We wish to find a function approximating $\Delta(t_i, u_j, p_j)$, which varies smoothly in time and sky coordinates. In this section, we outline a procedure for finding such a function using terms commonly applicable to astronomical analysis.

We describe sky mismatch using small shifts in right ascension and declination. For right ascension these shifts correspond to rotations about the Earth's equatorial axis. A shift in declination is not a rotation, but is a flow diverging from one pole and converging to the other. Because we use finite shift values, the declination flow is not defined in the pole vicinity, so a negligible region around each is excised from the input data. We define the following procedure:

\begin{enumerate}
\item Pick a set of signal arrival times ${\mathcal T}$. 

\item Construct a {\em  coarse} sky grid $G_t$ with minimum point separation of $\epsilon$ in spherical distance. Add to these a grid of points in a neighborhood $B_\odot$ around the Sun's position at each time $t\in {\mathcal T}$.

\item For every time $t\in \mathcal T$ and every point in the coarse sky grid $G_t$, compute the emission times $t_e \in T(t, G_t)$, where $T$ is a function returning a vector of emission times corresponding to each arrival time and source location.

\item Introduce a {\em displacement} grid $\Delta G$ of small sky rotations.

\item Compute emission times $T(t, G_{t,r})$ for each grid $G_{t,r}$ displaced by rotation $r \in \Delta G$.

\item Compute the difference $\Delta(t, G_t, r) \equiv T(t, G_{t,r})-T(t, G_{t,0})$ in emission times for each rotated and unrotated point at each time. This function varies smoothly in time and sky-direction. 

\item Define a function $\tilde\Delta(t, G_t, r)\equiv \sum_k a_k x_k$ with a set of coefficients $\{a_k\}$ for parameters $\{x_k\}$, and use least-squares fitting to compute $\{a_k\}$. Ideally, we would want to find $\{a_k\}$ such that $\mathtt{max}(|\tilde\Delta(t, G_t, r) - \Delta(t, G_t, r)|)$ is minimized, but the computational costs of such a search are too high. The parameters $\{x_k\}$ can be chosen for implementation convenience and are usually derived from easily computed (or pre-computed) quantities (see appendix \ref{sec:factors}).
\end{enumerate}

\section{Application example}
\label{sec:example}

\begin{table*}[htbp]
\caption{\label{tab:params} Input parameters used in fits}
\begin{tabular}{| l | l |}
\hline
\textbf{Parameter} & \textbf{Value}\\ \hline
${\mathcal T}$ & Every hour between $t_\mathrm{min}$ and $t_\mathrm{max}$\\ \hline
$t_\mathrm{min}$ & From start to end of O1, spaced every 200\,000 seconds\\ \hline
$t_\mathrm{max}$ & $t_\mathrm{min} + 250\,000$ seconds\\ \hline
$\epsilon$ & 0.1040524 rad\\ \hline
$\Delta G$ & All combinations of $\Delta\alpha$ and $\Delta\delta$\\ \hline
$G_{t,r}$ & Random subset of $G_t$ with $7.5\times 10^5$ points\\ \hline
$\Delta \alpha$ & $\{-0.01, -0.00667, -0.00333, 0, 0.00333, 0.00667, 0.01\}$\\ \hline
$\Delta \delta$ & $\{-0.01, -0.00667, -0.00333, 0, 0.00333, 0.00667, 0.01\}$\\ \hline
$\epsilon_\odot$ & 0.001 rad\\ \hline
$N_\odot$ & 5 \\ \hline
$\mathcal{S}(t)$ & Sun position at time $t$\\ \hline
$B_\odot$ & A grid of $N_\odot \times  N_\odot$ points centered on $\mathcal{S}(t)$, evenly spaced in $\alpha$ and $\delta$ with step $\epsilon_\odot$\\ \hline
\end{tabular}
\end{table*}

We now illustrate the algorithm with an application to real data. The regression factors are listed explicitly. We group them into several categories for ease of exposition. The grid parameters and other static inputs are listed in Table \ref{tab:params}. 

The categories group terms with similar composition:
\begin{itemize}
 \item Direction-independent terms depending on GPS time and shift in sky position
 \item Direction difference-independent terms depending on source sky position and GPS time
 \item Time-independent terms depending on source sky position and shift in position
\end{itemize}
In a practical implementation the direction-independent and time-independent terms could be precomputed, as the arrays needed to store them are relatively small. The direction difference-independent terms can be easily factorized into a product of precomputed arrays. All the terms are listed in appendix \ref{sec:factors}. 

The algorithm was applied to the time range covered by O1 data \cite{o1_data, losc, aligo} for 40 separate $250\,000$ second chunks, overlapping by $50\,000$ seconds each. 
As reference data, we use the tools included in the LIGO Analysis Library \cite{cws1}, which have been checked by comparison with the widely used radio astronomy timing package, TEMPO2 \cite{tempo,tempo2}. 

\begin{table}[htbp]
\caption{\label{tab:termRemoval} Term significance analysis. The max error column shows errors when the specified terms are omitted.}
\begin{tabular}{| l | l | l |}
\hline
\textbf{Term Group} & \textbf{Equation} & \textbf{Max Fit Error (s)}\\ \hline
2nd Order Sinusoids & \ref{eqn:trigSky2} & 3.4319967642 \\ \hline
1st Order Sinusoids & \ref{eqn:trigSky} & 0.4335595581 \\ \hline
$\Delta t$ & \ref{eqn:dt} & 0.0235629364 \\ \hline
$\Delta t^2$ & \ref{eqn:dt2} & 0.0010017764 \\ \hline
Sun Direction & \ref{eqn:sun} & 0.0002486640 \\ \hline
Sidereal Rotation & \ref{eqn:sidereal} & 0.0001820357 \\ \hline
Direction-difference & \ref{eqn:skyDiff} & 0.0001655650 \\ \hline
\end{tabular}
\end{table}

As a maximum acceptable error on timing, we used a 30-degree phase difference for a 2-kHz signal, or 42 $\mu$s, in order to
lose no more than $\sim$15\%\ SNR in all-sky CW searches reaching as high as 2 kHz. The SNR loss is less than $8$\% for a 1.5\,kHz signal.

An explicit fit formula for one of the chunks is listed in appendix \ref{sec:fit_example}.  This fit has the largest maximum error among the fits, 20.3 $\mu$s. The fit expression is a bilinear product of precomputed fit coefficients and monomials in $\Delta \alpha$, $\Delta \delta$ and $\Delta t$. In a practical implementation the grid of displacements, and thus monomial coefficients, is kept static inside the loop that computes $\Delta t$. The actual computation of $\Delta t$ easily vectorizes and takes few instructions on modern computers. Note that it is not necessary to keep the grid static with respect to all variables. 
For example, the grid can be static in $\Delta \delta$ and depend on $t$ and $\alpha$ --- the monomial grid recomputation cost will be amortized away.\\

Some fitting factors were necessary only because demodulation of high-frequency ($\approx 2$\,kHz) signals demands fine time resolution. Their influence on fit error is summarized in Table \ref{tab:termRemoval}. 
For lower-frequency signals the time resolution requirements are much looser, and some terms can be omitted from the fit.

\section{Results}
\label{sec:results}
The fits were tested using the following procedure.  We chose $16\times 8 = 128$ points on the sky, evenly spaced in right ascension and declination, to serve as patch centers.  For each patch, we shifted the central point by a random value in $[-0.01,0.01]$ for right ascension, and another for declination.  A total of 50 shifted points were generated for each patch. The even spacing of test points resulted in overcoverage at the poles, but this was tolerated in favor of code simplicity.

We divided the span of the first Advanced LIGO data run ($\sim$4 months), O1, into $200\,000$ second chunks, and took time points from each chunk at 30-minute intervals.  We obtained reference values for all points, then applied the fit model to each patch's points as a deflection from its center.  

\begin{figure}[htbp]
 \centering
 \includegraphics[width=0.45\textwidth]{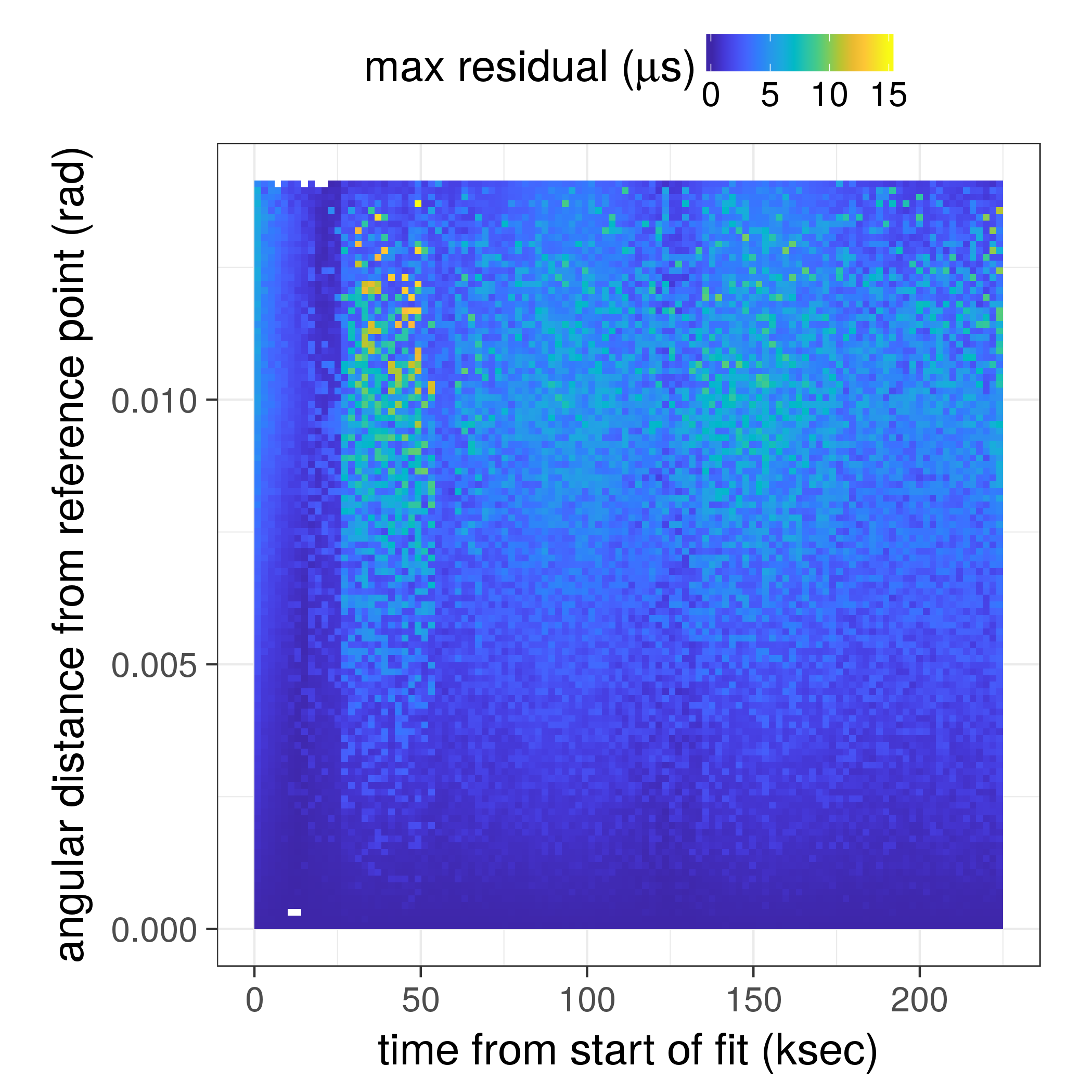}
 \caption{\label{fig:maxRes} Maximum absolute residual over all test patches. Angular distance is calculated as $\sqrt{(\Delta\alpha)^2 + (\Delta\delta)^2}$.}
\end{figure}

\begin{figure}[htbp]
 \centering
 \includegraphics[width=0.45\textwidth]{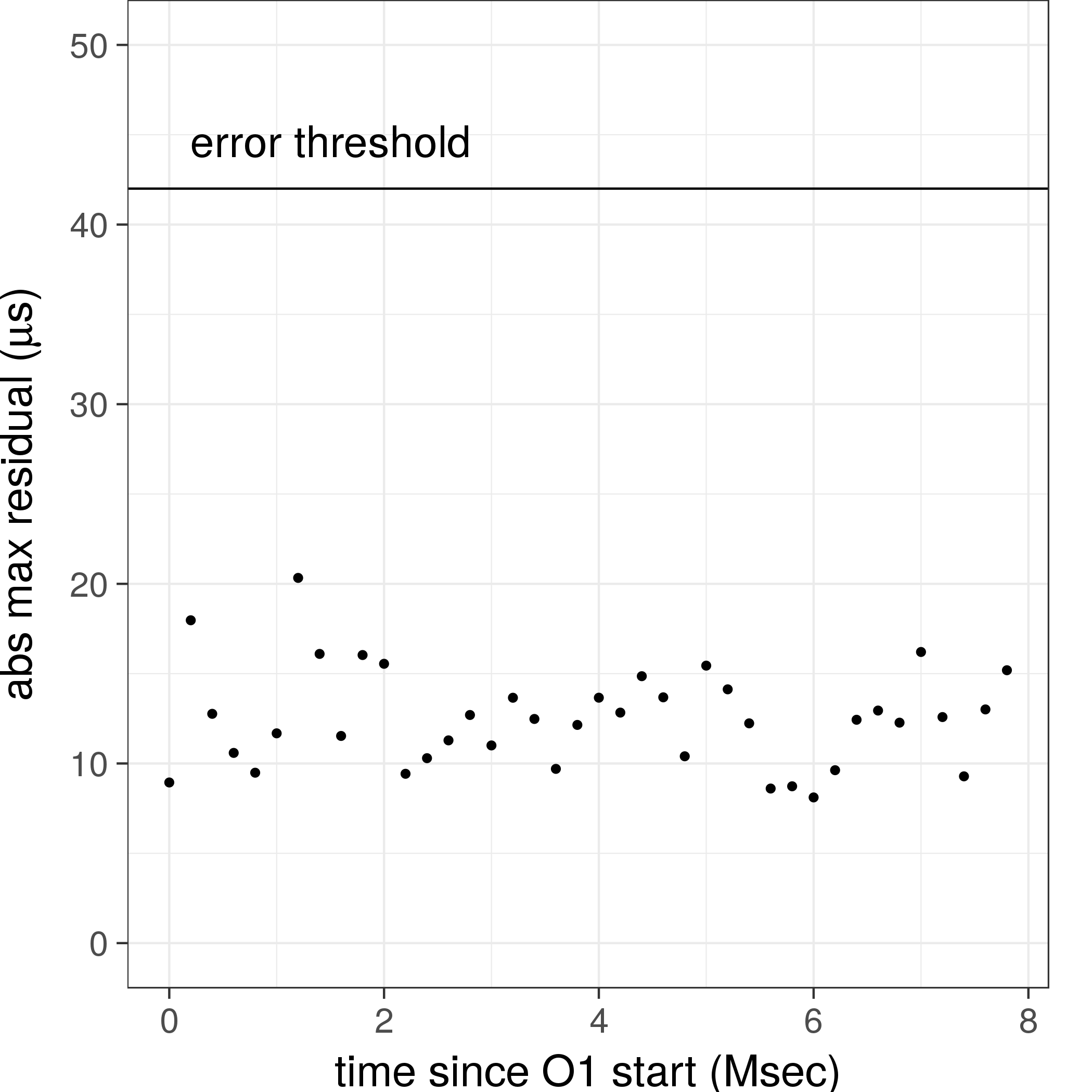}
 \caption{\label{fig:maxRes2} Maximum absolute residuals. Each point represents the fit for a 250 ksec interval. The fact that no periodicity is evident suggests that the fits are able to match the function behavior piecewise.}
\end{figure}

\begin{figure}[htbp]
 \centering
 \includegraphics[width=0.45\textwidth]{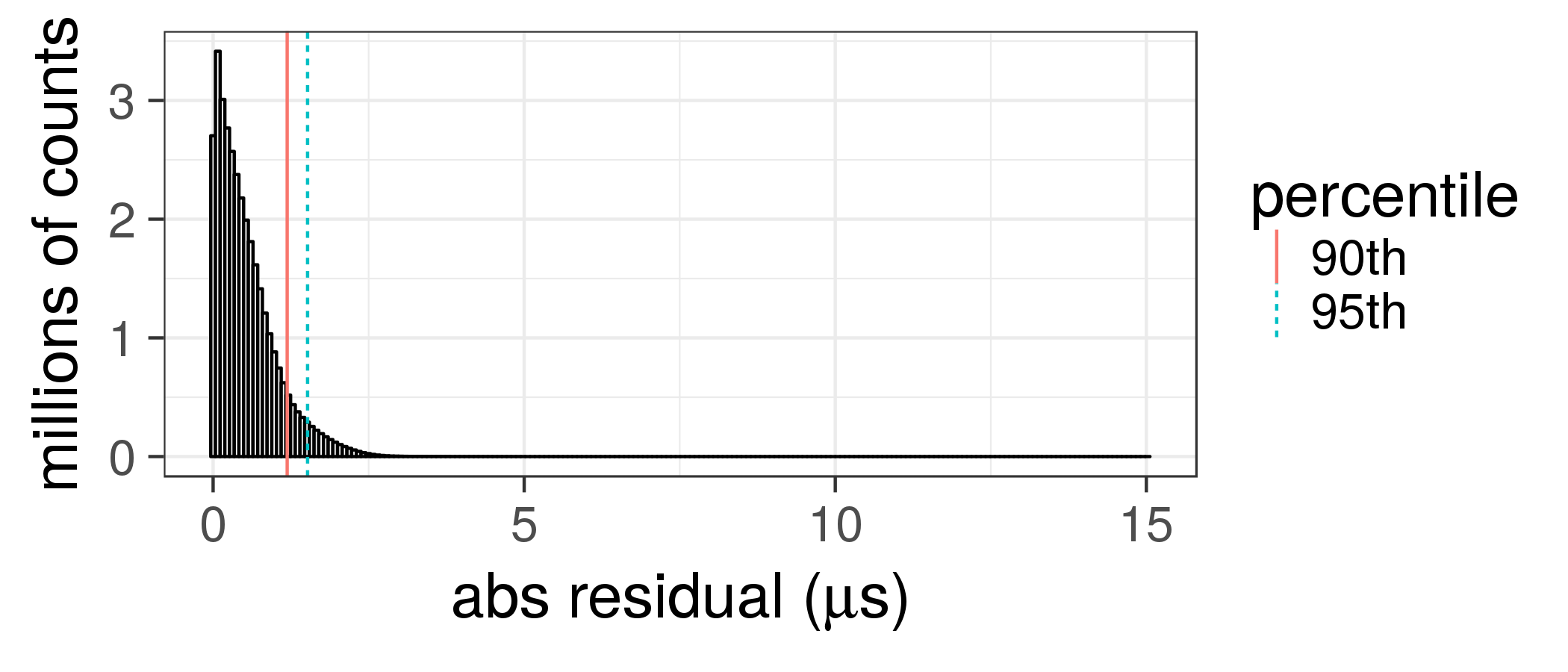}
 \caption{\label{fig:resDist} Distribution of residual magnitude. For large-timebase searches, only the bulk of the distribution matters, which is well below the error threshold.}
\end{figure}

A plot of the maximum absolute residual for each $\Delta t$ and $\Delta \phi$ is shown in Figure \ref{fig:maxRes}.  The maximum absolute residual for each reference time is shown in Figure \ref{fig:maxRes2}. All points fell below the error threshold. We also show a histogram of all errors in Figure \ref{fig:resDist}. The bulk of the errors are well below the threshold, and for a search of this length, any particular point would spend only a small fraction of time in a high-error region.

A prototype implementation of the method was optimized using SSE vector instructions, and a test was performed to compare the speed of the new method to the existing implementation in the {\tt LAL} library. The older SSE instruction set was chosen to demonstrate gains even on older computing hardware. We observed a 10x speedup relative to {\tt LAL} implementation. An implementation based on newer AVX512 instructions should show an additional 4x speedup. 
The prototype code computed sums $\sum_k a_kx_k$ for each template. In realistic search codes much of this computation can be reused, resulting in still larger speed improvements.

\section{Discussion}
\label{sec:conclusions}
The {\em Loosely Coherent} method of detecting signals analyzes sets of templates. For the set based on nearby sky locations it is important to understand the evolution of signal phases for nearby templates. The fit described in this paper explicitly demonstrates that relatively few parameters are needed to describe time arrival differences between nearby templates. While the benefits of reduced parameter count are clear for loosely coherent searches, the reduced count also has implications for conventional fully coherent searches.

It is well known that a mathematically optimal detection statistic consists of a linear filter followed by a power detector \cite{zubakov}. The linear filter is chosen to match expected signal properties and to reject noise outside of the signal bandwidth.
The presence of sky position difference terms in the example (appendix \ref{sec:fit_example}) shows that the sky position mismatch is equivalent to phase modulation of the incoming signal and the corrections span a 2-dimensional vector space. Thus for any search where sky position uncertainty requires multiple templates, the bandwidth of signals searched for is wider than the inverse of the integration time, and the fully coherent search is not the most efficient \cite{pfloose} detection statistic from both computational and detection viewpoints.

For example, if such a search uses one year's worth of data from a single interferometer, the maximal sensitivity is reached at 6 months integration time, or even earlier if parameters other than sky position are uncertain. For a search using many interferometers a fully coherent search can be more sensitive, but the gain in sensitivity is smaller than predicted from the increase of integration time alone.

This development provides an efficient method to compute emission time corrections, provides a basis for extension of the PowerFlux cache to longer integration times and lays the groundwork for future development of {\em Loosely Coherent} algorithms.

\section{Acknowledgements}
We thank members of the Continuous Waves Search Group of the LIGO Scientific Collaboration
and Virgo Collaboration for useful discussions.
This work was partly funded by National Science Foundation grant NSF PHY 1505932.

\appendix
\section{Regression factors}
\label{sec:factors}
In this section we list factors used in the regression fit.

\subsection{Definitions of Variables}
The sky position variables are defined as
\begin{equation}
\begin{array}{l}
e_1 = \cos(\delta)\cos(\alpha)\\
e_2 = \cos(\delta)\sin(\alpha)\\
e_3 = \sin(\delta)\\
\end{array}
\end{equation}
with $\alpha \in [-\pi, \pi]$ and $\delta \in [-\tfrac{\pi}{2} + 0.01, \tfrac{\pi}{2} - 0.01]$. The adjustment by $0.01$ radians prevents flow over the poles, which would lead to ambiguous right ascension. The change in $e_i$ for a shift in right ascension $\Delta\alpha$, and in declination $\Delta\delta$ can be approximated via Taylor expansion:
\begin{equation}
\begin{array}{l}
\Delta e_1 = (-\tfrac{1}{2} \Delta\alpha^2 \cos\alpha \cos\delta-\tfrac{1}{2} \Delta\delta^2 \cos\alpha \cos\delta\\
\qquad+\tfrac{1}{4} \Delta\alpha^2 \Delta\delta^2 \cos\alpha \cos\delta- \Delta\alpha \cos\delta \sin\alpha\\
\qquad+\tfrac{1}{2} \Delta\alpha \Delta\delta^2 \cos\delta \sin\alpha-\Delta\delta \cos\alpha \sin\delta\\
\qquad+\tfrac{1}{2} \Delta\alpha^2 \Delta\delta \cos\alpha \sin\delta+\Delta\alpha \Delta\delta \sin\alpha \sin\delta)\\
\Delta e_2 = (\Delta\alpha \cos\alpha \cos\delta-\tfrac{1}{2} \Delta\alpha \Delta\delta^2 \cos\alpha \cos\delta\\
\qquad-\tfrac{1}{2} \Delta\alpha^2 \cos\delta \sin\alpha-\tfrac{1}{2} \Delta\delta^2 \cos\delta \sin\alpha\\
\qquad+\tfrac{1}{4} \Delta\alpha^2 \Delta\delta^2 \cos\delta \sin\alpha-\Delta\alpha \Delta\delta \cos\alpha \sin\delta\\
\qquad-\Delta\delta \sin\alpha \sin\delta+\tfrac{1}{2} \Delta\alpha^2 \Delta\delta \sin\alpha \sin\delta)\\
\Delta e_3 = (\Delta\delta \cos\delta-\tfrac{1}{2} \Delta\delta^2 \sin\delta)\\
\end{array}
\end{equation}
For each time point we define,
\begin{equation}
\begin{array}{l l}
\vec{\mathcal{S}} & \textrm{Vector pointing from Sun\ to Earth}\\
\vec{v} & \textrm{Detector velocity vector}\\
\Delta t & \textrm{Time since reference point}\\
\Omega_{\oplus} & 2\pi/\textrm{sidereal day}.
\end{array}
\end{equation}
We also define an array of the sin/cos of the reference point's right ascension and declination:
\begin{equation}
\mathbf{z} = \{\sin\alpha, \sin\delta, \cos\alpha, \cos\delta\}
\end{equation}
and the second-order terms, excepting $\sin^2$ terms because they can be expressed as $1-\cos^2$:
\begin{equation}
\begin{array}{l}
\mathbf{z'} = \{\cos^2\alpha, \cos^2\delta, \sin\alpha \sin\delta, \sin\alpha \cos\alpha, \\
\qquad \sin\alpha \cos\delta, \sin\delta \cos\alpha, \sin\delta \cos\delta, \cos\alpha \cos\delta\}
\end{array}
\end{equation}

\subsection{Direction-independent terms}
The following terms are constant in sky-direction, and can be precomputed for every GPS time, and direction difference.

\begin{equation}
\label{eqn:skyDiff}
\sum_{i} a_{1,i} \Delta e_i
\end{equation}

\subsection{Difference-independent terms}
The following terms are constant in direction-difference.
\begin{equation}
\label{eqn:sidereal}
\sum_{i} a_{2,i} \sin(\Omega_{\oplus} \Delta t) z'_i + b_{2,i} \sin(\Omega_{\oplus} \Delta t) z'_i
\end{equation}
\begin{equation}
\label{eqn:dt}
a_{3,1} \Delta t \cos\delta + a_{3,2} \Delta t^2 \cos\delta + a_{3,3} \Delta t \cos^2\delta
\end{equation}
\begin{equation}
\label{eqn:dt2}
\sum_{i} a_{4,i} \Delta t^2 z'_i
\end{equation}
\begin{equation}
\label{eqn:sun}
\sum_{i} a_{5,i} \mathcal{S}_i e_i
\end{equation}

\subsection{Time-independent terms}
The following terms vary only in sky-direction.
\begin{equation}
\label{eqn:trigSky}
\sum_{i} a_{6,i} z_i
\end{equation}
\begin{equation}
\label{eqn:trigSky2}
\sum_{i} a_{7,i} z'_i
\end{equation}

Each of the terms in equations \ref{eqn:skyDiff}-\ref{eqn:trigSky2} is multiplied by $\Delta \alpha, \Delta \delta, \Delta \alpha^2, \Delta \delta^2, \Delta \alpha\Delta \delta$.  In addition, we include direction-time differential terms
\begin{equation}
\sum_{i} a_{8,i} \Delta t \Delta e_i
\end{equation}
\noindent without $\Delta \alpha/\Delta \delta$ factors.  Each term goes to zero when the rotation angle goes to zero.  Note that Sun-Earth and detector velocity vectors are those for the saved points. In each term, any parts greater than order 3 in $\Delta \alpha$ and $\Delta \delta$ are removed. The effects of removing sets of terms are shown in Table \ref{tab:termRemoval}.

\begin{widetext}

\section{Example Fit}
\label{sec:fit_example}

As an example, we list below the resulting formula from a fit for GPS time 1127833121. The expression is a bilinear product between precomputed fit coefficients and monomials in $\Delta \alpha$, $\Delta \delta$ and $\Delta t$. Only significant terms are shown.
 This fit has the largest maximum error among the fits, 20.3 $\mu$s.
 
\begin{samepage}
\begin{align*}
\Delta T =\ &\ 
\Large[-1.69 \Large] 10^{-5} \mathbf{\Delta t} \mathbf{\Delta e_1} + \Large[8.98 \Large] 10^{-5} \mathbf{\Delta t} \mathbf{\Delta e_2} + \Large[-495 e_2+71.2 e_1\Large] \mathbf{\Delta \alpha} +\\
&  \Large[30.8 \cos(\delta)-71.2 e_3 \sin(\alpha)-495 e_3 \cos(\alpha)\Large] \mathbf{\Delta \delta} +\Large[3.89 \cos(\delta)\Large] 10^{-5} \mathbf{\Delta \delta} \mathbf{\Delta t} + \Large[8.14 e_2+83 e_1\Large] \mathbf{\Delta \alpha}^2 +\\
&  \Large[0.0132 e_2-0.00634 e_1\Large] \sin(\Omega_{\oplus} \mathbf{\Delta t}) \mathbf{\Delta \alpha} + \Large[0.00644 e_2+0.0132 e_1\Large] \cos(\Omega_{\oplus} \mathbf{\Delta t}) \mathbf{\Delta \alpha} +\\
&  \Large[-35.6 e_2-248 e_1-0.5 e_3 \mathcal{S}_3\Large] \mathbf{\Delta \delta}^2 + \Large[0.00634 e_3 \sin(\alpha)+0.0132 e_3 \cos(\alpha)\Large] \sin(\Omega_{\oplus} \mathbf{\Delta t}) \mathbf{\Delta \delta} +\\
&  \Large[0.0972 e_2-0.0157 e_1\Large] 10^{-10} \mathbf{\Delta \alpha} \mathbf{\Delta t}^2 + \Large[-0.0132 e_3 \sin(\alpha)+0.00644 e_3 \cos(\alpha)\Large] \cos(\Omega_{\oplus} \mathbf{\Delta t}) \mathbf{\Delta \delta} +\\
&  \Large[-0.00686 \cos(\delta)+0.0157 e_3 \sin(\alpha)+0.0972 e_3 \cos(\alpha)\Large] 10^{-10} \mathbf{\Delta \delta} \mathbf{\Delta t}^2 + \Large[43.7 \Large] \mathbf{\Delta \alpha} \mathbf{\Delta e_1} + \Large[-331 \Large] \mathbf{\Delta \alpha} \mathbf{\Delta e_2} +\\
&  \Large[83.2 \Large] \mathbf{\Delta \alpha}^2 \mathbf{\Delta e_1} + \Large[-82.2 \Large] \mathbf{\Delta \delta}^2 \mathbf{\Delta e_1} + \Large[165 e_3 \sin(\alpha)-27.4 e_3 \cos(\alpha)\Large] \mathbf{\Delta \alpha} \mathbf{\Delta \delta} + \Large[-0.0972 e_3 \sin(\alpha)\Large] 10^{-10} \mathbf{\Delta \alpha} \mathbf{\Delta \delta} \mathbf{\Delta t}^2
\end{align*}
\end{samepage}
\end{widetext}

\bibliography{EarthFit,PrevSearches}

\begin{thebibliography}{10}

\bibitem{CasAPaper}
J.~Abadie et~al.
\newblock {First search for gravitational waves from the youngest known neutron
  star}.
\newblock {\em Astrophys.J.}, 722:1504--1513, 2010.

\bibitem{CrabPaper}
B.~Abbott et~al.
\newblock {Beating the spin-down limit on gravitational wave emission from the
  Crab pulsar}.
\newblock {\em Astrophys.J.}, 683:L45--L50, 2008.

\bibitem{EarlyS5PowerFluxPaper}
B.~Abbott et~al.
\newblock {All-sky LIGO Search for Periodic Gravitational Waves in the Early S5
  Data}.
\newblock {\em Phys.Rev.Lett.}, 102:111102, 2009.

\bibitem{FullS5PowerFluxPaper}
J.~Abadie et~al.
\newblock {All-sky Search for Periodic Gravitational Waves in the Full S5 LIGO
  Data}.
\newblock {\em Phys.Rev.}, D85:022001, 2012.

\bibitem{houghS2}
B.~Abbott et~al.
\newblock {First all-sky upper limits from LIGO on the strength of periodic
  gravitational waves using the Hough transform}.
\newblock {\em Phys.Rev.}, D72:102004, 2005.

\bibitem{LSC:05-S2TD}
B.~Abbott et~al.
\newblock {Limits on gravitational wave emission from selected pulsars using
  LIGO data}.
\newblock {\em Phys.Rev.Lett.}, 94:181103, 2005.

\bibitem{LSC:07-S34TD}
B.~Abbott et~al.
\newblock {Upper limits on gravitational wave emission from 78 radio pulsars}.
\newblock {\em Phys.Rev.}, D76:042001, 2007.

\bibitem{O1knownPulsars}
B.~P. Abbott et~al.
\newblock First search for gravitational waves from known pulsars with advanced
  {LIGO}.
\newblock {\em The Astrophysical Journal}, 839(1):12, 2017.

\bibitem{orionspur}
J.~Aasi et~al.
\newblock A search of the orion spur for continuous gravitational waves using a
  "loosely coherent" algorithm on data from ligo interferometers.
\newblock {\em Physical Review D}, 93, 10 2015.

\bibitem{PhysRevD.93.042007}
J.~Aasi et~al.
\newblock First low frequency all-sky search for continuous gravitational wave
  signals.
\newblock {\em Phys. Rev. D}, 93:042007, Feb 2016.

\bibitem{PhysRevD.94.064061}
A.~Singh, M.~A. Papa, H.-B. Eggenstein, S.~Zhu, H.~Pletsch, B.~Allen, O.~Bock,
  B.~Maschenchalk, R.~Prix, and X.~Siemens.
\newblock Results of an all-sky high-frequency einstein@home search for
  continuous gravitational waves in ligo's fifth science run.
\newblock {\em Phys. Rev. D}, 94:064061, Sep 2016.

\bibitem{PhysRevD.95.082005}
B.~P. Abbott et~al.
\newblock Search for continuous gravitational waves from neutron stars in
  globular cluster ngc 6544.
\newblock {\em Phys. Rev. D}, 95:082005, Apr 2017.

\bibitem{PhysRevD.96.062002}
B.~P. Abbott et~al.
\newblock All-sky search for periodic gravitational waves in the o1 ligo data.
\newblock {\em Phys. Rev. D}, 96:062002, Sep 2017.

\bibitem{PhysRevD.96.122004}
B.~P. Abbott et~al.
\newblock First low-frequency einstein@home all-sky search for continuous
  gravitational waves in advanced ligo data.
\newblock {\em Phys. Rev. D}, 96:122004, Dec 2017.

\bibitem{PhysRevLett.118.121102}
B.~P. Abbott et~al.
\newblock Directional limits on persistent gravitational waves from advanced
  ligo's first observing run.
\newblock {\em Phys. Rev. Lett.}, 118:121102, Mar 2017.

\bibitem{S1PulPaper}
B.~Abbott et~al.
\newblock {Setting upper limits on the strength of periodic gravitational waves
  using the first science data from the GEO 600 and LIGO detectors}.
\newblock {\em Phys.Rev.}, D69:082004, 2004.

\bibitem{S2Fstat}
B.~Abbott et~al.
\newblock {Coherent searches for periodic gravitational waves from unknown
  isolated sources and Scorpius X-1: Results from the second LIGO science run}.
\newblock {\em Phys.Rev.}, D76:082001, 2007.

\bibitem{S4EaHpaper}
B.~Abbott et~al.
\newblock {The Einstein@Home search for periodic gravitational waves in LIGO S4
  data}.
\newblock {\em Phys.Rev.}, D79:022001, 2009.

\bibitem{S4IncoherentPaper}
B.~Abbott et~al.
\newblock {All-sky search for periodic gravitational waves in LIGO S4 data}.
\newblock {\em Phys.Rev.}, D77:022001, 2008.

\bibitem{S5Hough}
J.~Aasi et~al.
\newblock {Application of a Hough search for continuous gravitational waves on
  data from the fifth LIGO science run}.
\newblock {\em Class.Quant.Grav.}, 31:085014, 2014.

\bibitem{S5R1EaHpaper}
B.~Abbott et~al.
\newblock {Einstein at Home search for periodic gravitational waves in early S5
  LIGO data}.
\newblock {\em Phys.Rev.}, D80:042003, 2009.

\bibitem{S5R5EaHpaper}
J.~Aasi et~al.
\newblock {Einstein@Home all-sky search for periodic gravitational waves in
  LIGO S5 data}.
\newblock {\em Phys.Rev.}, D87(4):042001, 2013.

\bibitem{S5Targeted}
B.~Abbott et~al.
\newblock {Searches for gravitational waves from known pulsars with S5 LIGO
  data}.
\newblock {\em Astrophys.J.}, 713:671--685, 2010.

\bibitem{S6Bucket}
B.~P. Abbott, et~al.
\newblock Results of the deepest all-sky survey for continuous gravitational
  waves on ligo s6 data running on the einstein@home volunteer distributed
  computing project.
\newblock {\em Phys. Rev. D}, 94:102002, Nov 2016.

\bibitem{S6PostCasAPaper}
J.~Aasi, et~al.
\newblock {Searches for continuous gravitational waves from nine young
  supernova remnants}.
\newblock {\em The Astrophysical Journal}, 813(1):39, oct 2015.

\bibitem{S6PowerFluxPaper}
B.~P. Abbott, et~al.
\newblock Comprehensive all-sky search for periodic gravitational waves in the
  sixth science run ligo data.
\newblock {\em Phys. Rev. D}, 94:042002, Aug 2016.

\bibitem{S6SpotlightPaper}
J.~{Aasi} et~al.
\newblock {Search of the Orion spur for continuous gravitational waves using a
  loosely coherent algorithm on data from LIGO interferometers}.
\newblock {\em \prd}, 93(4):042006, February 2016.

\bibitem{VSR1fstat}
J.~Aasi et~al.
\newblock {Implementation of an F-statistic all-sky search for continuous
  gravitational waves in Virgo VSR1 data}.
\newblock {\em Class.Quant.Grav.}, 31:165014, 2014.

\bibitem{ROMpaper}
M.~Pitkin, S.~Doolan, L.~McMenamin, and K.~Wette.
\newblock Reduced order modelling in searches for continuous gravitational
  waves -- i. barycentring time delays.
\newblock {\em Monthly Notices of the Royal Astronomical Society},
  476(4):4510--4519, 2018.

\bibitem{pfloose}
V.~Dergachev.
\newblock On blind searches for noise dominated signals: a loosely coherent
  approach.
\newblock {\em Class. Quantum Grav.}, 27(205017), 2010.

\bibitem{pfloose2}
V.~Dergachev.
\newblock Loosely coherent searches for sets of well-modeled signals.
\newblock {\em Phys.\ Rev.\ D}, 85(062003), 2012.

\bibitem{loosely_coherent3}
V.~Dergachev.
\newblock Efficient loosely coherent searches for medium scale coherence
  lengths.
\newblock {\em arXiv:1807.02351}, 2018.

\bibitem{s5}
J.~Abadie, et~al.
\newblock All-sky search for periodic gravitational waves in the full {S5}
  {LIGO} data.
\newblock {\em Physical Review D}, 85(2):022001, 2012.

\bibitem{s6}
B.~P. Abbott, et~al.
\newblock Comprehensive all-sky search for periodic gravitational waves in the
  sixth science run {LIGO} data.
\newblock {\em Physical Review D}, 94(4):042002, 2016.

\bibitem{o1allsky}
B.~P. Abbott et~al.
\newblock All-sky search for periodic gravitational waves in the {O1} {LIGO}
  data.
\newblock {\em Phys. Rev. D}, 96:062002, Sep 2017.

\bibitem{tempo2}
R.~T. Edwards, G.~B. Hobbs, and R.~N. Manchester.
\newblock tempo2, a new pulsar timing package -- ii. the timing model and
  precision estimates.
\newblock {\em Monthly Notices of the Royal Astronomical Society},
  372(4):1549--1574, 2006.

\bibitem{kolmogorov}
A.~N. Kolmogorov.
\newblock {\em Selected Works of A. N. Kolmogorov. Mathematics and Its
  Applications (Soviet Series)}, volume~25, chapter On the representation of
  continuous functions of several variables by superpositions of continuous
  functions of a smaller number of variables, pages 378--382.
\newblock Springer, Dordrecht, 1991.

\bibitem{arnold}
A.~B. Givental, B.~A. Khesin, J.~E. Marsden, A.~N. Varchenko, V.~A. Vassiliev,
  O.~Y. Viro, and V.~M. Zakalyukin, editors.
\newblock {\em On functions of three variables}, pages 5--8.
\newblock Springer Berlin Heidelberg, Berlin, Heidelberg, 2009.

\bibitem{saturated}
S.~E. Field, C.~R. Galley, J.~S. Hesthaven, J.~Kaye, and M.~Tiglio.
\newblock Fast prediction and evaluation of gravitational waveforms using
  surrogate models.
\newblock {\em Phys. Rev. X}, 4:031006, Jul 2014.

\bibitem{o1_data}
Ligo open science center.
\newblock {\em \url{https://losc.ligo.org}}.

\bibitem{losc}
M.~Vallisneri et~al.
\newblock The ligo open science center.
\newblock {\em proceedings of the 10th LISA Symposium, University of Florida,
  Gainesville,arxiv:1410.4839}, 2014.

\bibitem{aligo}
J.~Aasi, et~al.
\newblock Advanced {LIGO}.
\newblock {\em Classical and Quantum Gravity}, 32(7):074001, mar 2015.

\bibitem{cws1}
B.~Abbott et~al.
\newblock Setting upper limits on the strength of periodic gravitational waves
  from psr $\mathrm{J}1939+2134$ using the first science data from the geo 600
  and ligo detectors.
\newblock {\em Phys. Rev. D}, 69:082004, Apr 2004.

\bibitem{tempo}
R.~T. Edwards, G.~B. Hobbs, and R.~N. Manchester.
\newblock \url{www.atnf.csiro.au/research/pulsar/tempo}.

\bibitem{zubakov}
L.~Wainstein and V.~Zubakov.
\newblock {\em Extraction of Signals from Noise}.
\newblock Dover Publications Inc, 1971.

\end{thebibliography}

\end{document}